# Fresnel diffraction of cylindrical and spherical wavefronts from a phase plate

Masoud Ghoorchi-Beygi and Masoomeh Dashtdar*

*Department of physics, Shahid Beheshti University, G. C., Evin, Tehran 19839, Iran*

**Corresponding author Email: m-dashtdar@sbu.ac.ir*

**Abstract:** In the last two decades, Fresnel diffraction (FD) of a plane wave from phase steps has been systematically studied and applied for precise measurements of light wavelength, and height and refractive index of the step. In this study we formulate FD of cylindrical and spherical wavefronts from a transparent step in transmission mode. It is shown that, the intensity distribution is a periodic function along the lines parallel to the plate edge. The phase distribution in this direction is a linearly varying function of the position squared, with a slope depends on the light wavelength, the plate thickness and refractive index, and the radius of wavefront curvature (RWC) on the observation plane. Therefore, it has significant potential in optical metrology, by a single-shot recording. The diffraction patterns are simulated and experimentally verified. Also, the RWC and displacement are determined as examples of applications in the experimental part of the report.

## 1. INTRODUCTION

The familiar Fresnel diffraction (FD) occurs when the passage of a wave is partially obstructed by an aperture. In this case, the amplitude of the wave experiences a sharp change at the boundary regions. FD of matter, sound, and electromagnetic waves from different kinds of aperture shapes has been widely studied [1-7] and applied in various fields [8-11], especially in amplitude and phase retrieval in optical devices and systems [12-17]. FD occurs also as a wave travel through a medium with a varying phase (i.e. phase object). Measurements of thermal lens and nonlinear refractive index of liquids are examples of its application [18,19]. In addition, it is observed that a sharp change in the phase of a coherent light beam causes significant FD. A sharp change in phase occurs when the beam reflects from a physical step or transmits through the boundary region of a phase plate. The FD of a plane wave from a phase step has been studied in detail [20-23]. The visibility of the diffraction fringes varies considerably by changing the light incident angle that provides precise measurements of refractive indices of solids and liquids [24-26], film thickness [27,28], wavelength [29,30], and focal length of imaging systems [31,32].

In this paper, we present the formulation of the FD of cylindrical and spherical waves from a phase plate, which is an important topic in physical optics. Since, the phase difference of the spherical (or cylindrical) wave varies continuously along the phase plate edge, it can be applied by a single-shot recording in optical metrology, such as the phase distribution, radius of wavefront curvature (RWC), wavelength, thickness and refractive index of the step, and displacement. The intensity distribution on diffraction patterns are simulated and experimentally verified. The theoretical results show that the phase distribution on the FD pattern along the lines parallel to the plate edge is a linear function of the position squared, and is independent to the plate position. It is confirmed by the experimental results. Also, the RWC and displacement are measured as application examples.

Determination of the phase distribution and the RWC are important in many application areas, for example, quantitative phase imaging [33,34], adaptive optics [35], and optical surface profilometry [36,37]. The most common methods for reconstruction of phase distribution are Shack-Hartmann sensor [38], and interferometry [39]. The FD based methods are easy to apply and do not require complex data processing. In addition, the measuring setups have very low optical and mechanical noises compared with interferometric methods.

## 2. THEORETICAL APPROACH

### 2.1. Fresnel diffraction of cylindrical wavefront from a phase plate

A sketch of FD of a line source from the edge of a transparent plate is shown in Fig. 1(a). Figure 1(b) illustrates a top view of Fig. 1(a). A monochromatic cylindrical wave ($\Sigma$) with symmetry axis perpendicular to the page ($xz$-plane) at point S strikes a transparent plane-parallel plate in the boundary area of the upper edge. The plate is placed at distance $z_0$ from the point S, so that its edge is perpendicular to the symmetry axis.

By transmitting through the plate, the beam experiences a sharp change in phase at the boundary of the plate. The phase distribution on the $x'y'$-plane just after the plate is obtained by considering two sets of rays that reach the $x'y'$-plane at same $x'$; one propagates above the plate, without deflection (for $y' > 0$), and the other passing through the plate at incident angle $\theta$ (for $y' < 0$):

$$\varphi(x',y') = k\begin{cases}\sqrt{z_0^2 + z_0^2\tan^2\theta} + n\dfrac{h}{cos\beta} & y' < 0\\ \sqrt{x'^2 + (z_0+h)^2} & y' > 0\end{cases}, \quad (1)$$

where $k = 2\pi/\lambda$ is the wavenumber, $h$ and $n$ are the plate thickness and refractive index, and $\beta$ is the refraction angle.

According to geometry in Fig. 1(b),

$$\tan\theta = (\frac{x'+d}{z_0+h}), \quad (2)$$

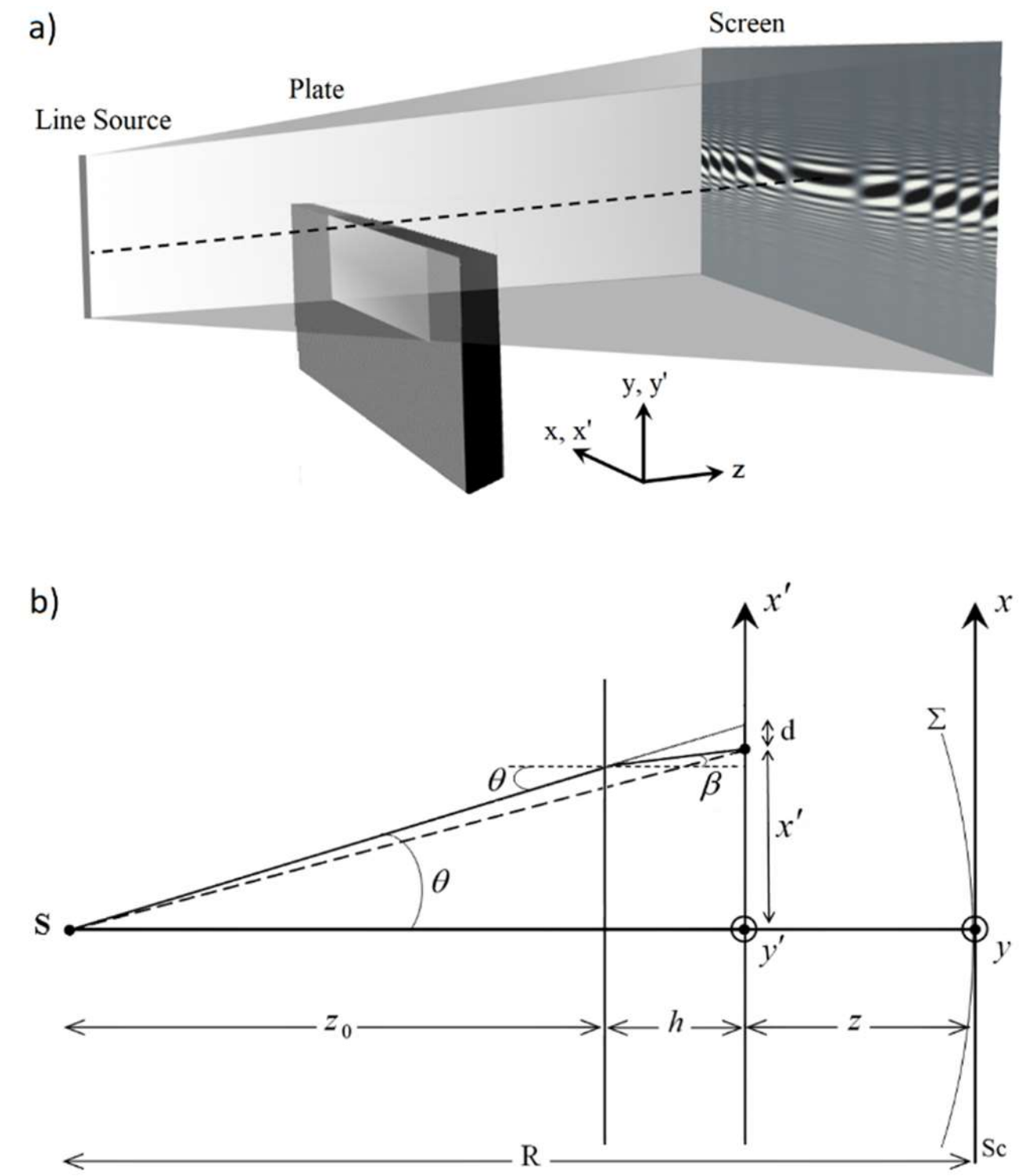

**Fig. 1.** (a) A sketch of diffraction of a line source of a monochromatic light from the edge of a transparent plate. (b) The geometry used to calculate the optical Fresnel diffraction of cylindrical and spherical waves from a phase plate.

where $d = h(\tan\theta - \tan\beta)$ is the lateral displacement of the rays passing through the plate. By applying Snell's law and small angle approximations, the phase distribution in Fresnel approximation takes the following form:

$$\varphi(x', y') = k\begin{cases} \dfrac{n}{2(nz_0+h)}x'^2 + z_0 + nh & y' < 0 \\ \dfrac{1}{2(z_0+h)}x'^2 + z_0 + h & y' > 0 \end{cases}. \tag{3}$$

The complex amplitude on the observation plane ($xy$-plane) at distance $z$ from the $x'y'$–plane can be given by applying the Fresnel-Kirchhoff integral as following form [40]:

$$\psi(x,y) = KA \times\left\{ t\int_{-\infty}^{0}\int_{-\infty}^{\infty} e^{ik\left(\frac{n}{2(nz_0+h)}x'^2+nh+z_0\right)} e^{\frac{ik}{2z}\left((x-x')^2+(y-y')^2\right)}dx'\,dy' + \int_{0}^{\infty}\int_{-\infty}^{\infty} e^{ik\left(\frac{1}{2(z_0+h)}x'^2+h+z_0\right)} e^{\frac{ik}{2z}\left((x-x')^2+(y-y')^2\right)}dx'\,dy' \right\}, \tag{4}$$

where $K$, $A$, and $t$ stand for propagation factor, incident amplitude and amplitude transmission coefficient. Evaluating the integrals lead to:

$$\begin{aligned}\psi(x,y) = {} & \frac{KA}{2}\lambda\sqrt{z}(1+i) \\ & \times\left\{ t\left[\frac{1}{2}+\frac{i}{2}+C(\alpha)+i\mathrm{S}(\alpha)\right]\left(\frac{n}{nz_0+h}+\frac{1}{z}\right)^{-\frac{1}{2}} \right. \\ & \times e^{i\frac{k}{2}\left(\frac{n}{nz_0+h+nz}x^2+2(\mathrm{n}-1)\mathrm{h}\right)} + \left[\frac{1}{2}+\frac{i}{2}-C(\alpha)-i\mathrm{S}(\alpha)\right] \\ & \left. \times\left(\frac{1}{z_0+h}+\frac{1}{z}\right)^{-\frac{1}{2}} e^{i\frac{k}{2}\left(\frac{1}{z_0+h+z}x^2\right)} \right\},\end{aligned} \tag{5}$$

where $C(\alpha)$ and $S(\alpha)$ are Fresnel integrals and $\alpha$ defined as $\alpha = y\sqrt{2/\lambda z}$.

To obtain the normalized intensity at $xy$-plane, we multiply Eq. (5) by its complex conjugate and divide the latter by the intensity in the absence of the plate which, after some manipulations, leads to

$$\begin{aligned} I_n(x,y) = {} & \frac{Dt^2+1}{4} + \left(\frac{Dt^2-1}{2}\right)\left[\mathrm{C}(\alpha)+\mathrm{S}(\alpha)\right] \\ & + \left(\frac{Dt^2+1}{2}\right)\left[\mathrm{C}^2(\alpha)+\mathrm{S}^2(\alpha)\right] \\ & + t\sqrt{D}\left\{\left[\frac{1}{2}-\mathrm{C}^2(\alpha)-\mathrm{S}^2(\alpha)\right]cos\phi(x) \right. \\ & \left. + \left[\mathrm{C}(\alpha)-\mathrm{S}(\alpha)\right]sin\phi(x)\right\}, \end{aligned} \tag{6}$$

where

$$D = \frac{R(\mathrm{n}z_0+\mathrm{h})}{(\mathrm{nR}-\mathrm{nh}+\mathrm{h})(z_0+h)}, \tag{7}$$

$$\phi(x) = \frac{k}{2}\left(\frac{n}{nR-nh+h}-\frac{1}{R}\right)x^2 + \mathrm{k}(\mathrm{n}-1)\mathrm{h}, \tag{8}$$

and $R = z_0 + h + z$ is the RWC on the observation plane.

The normalized intensity specified by Eq. (6) is important and interesting in optical metrology. It is a periodic function along the lines parallel to the plate edge ($x$ direction), expressly, as $\phi\,(x)$ changes by $2\pi$ the intensities are repeated along perpendicular to the plate edge ($y$ direction). It provides a single-shot phase shifting technique for optical measurements. In addition, according to Eq. (8) $\phi$ is a linearly varying function of $x^2$ with a slope depends on the light wavelength, the plate thickness and refractive index, and the RWC on the observation plane. Therefore, the parameters can be obtained by determining the phase distribution.

It is worth mentioning that the slope is independent to the plate position that provides a straight forward approach to measure the RWC. It can be achieved by two methods. In the first method, the RWC is obtained by fitting Eq. (8) to the experimental plot of phase change versus $x$ (or $x^2$). In the second method, the RWC is obtained

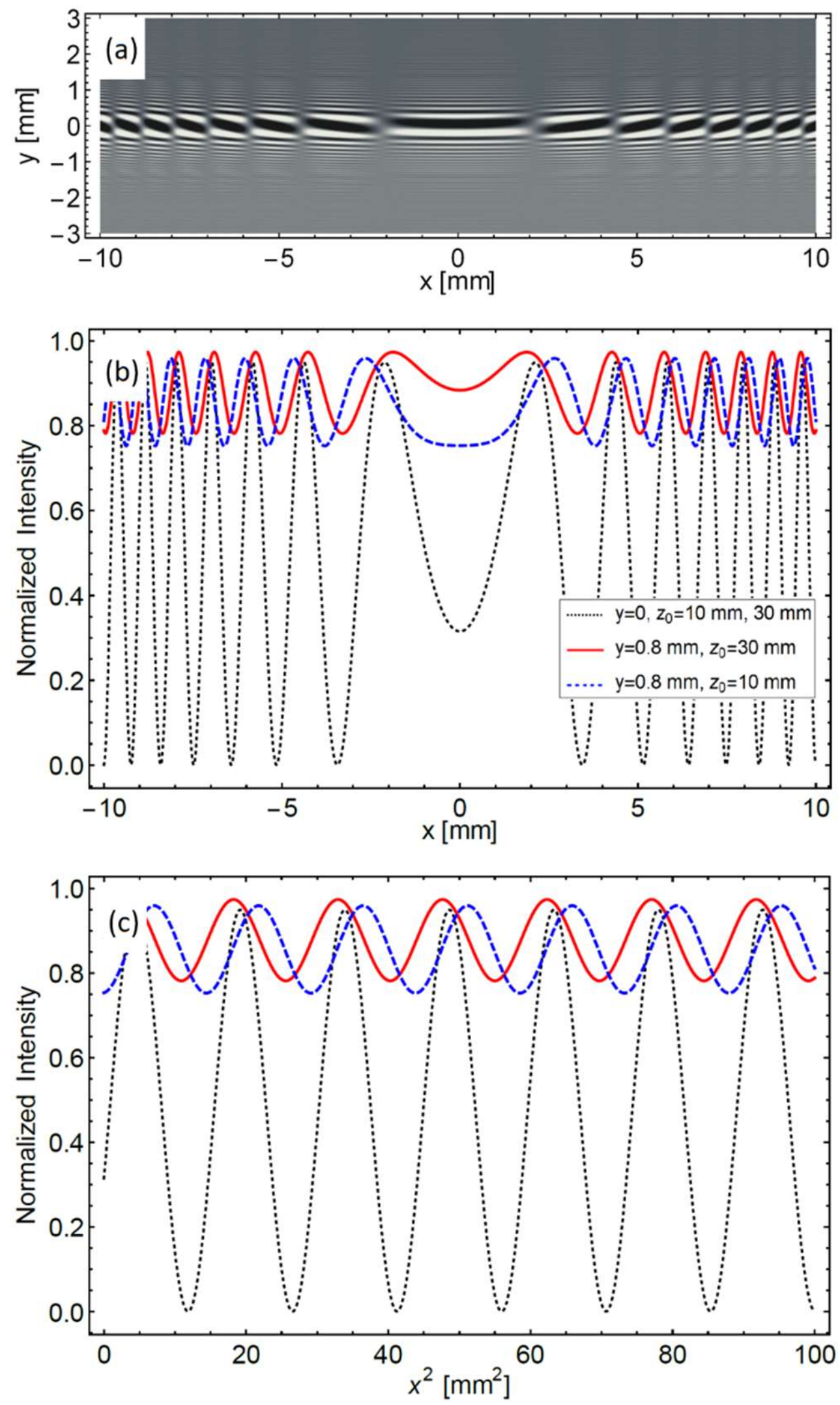

**Fig. 2.** (a) The simulated Fresnel diffraction pattern of cylindrical wave from a phase plate with $t = 1, h = 5$ mm and $n = 1.5$, for $\lambda = 632.8$ nm, $R = 140$ mm, and $z_0 = 10$ mm. (b) The normalized intensity profiles in the direction parallel to the plate edge for different values of $y$ and $z_0$, and (c) the corresponding normalized intensity versus $x^2$.

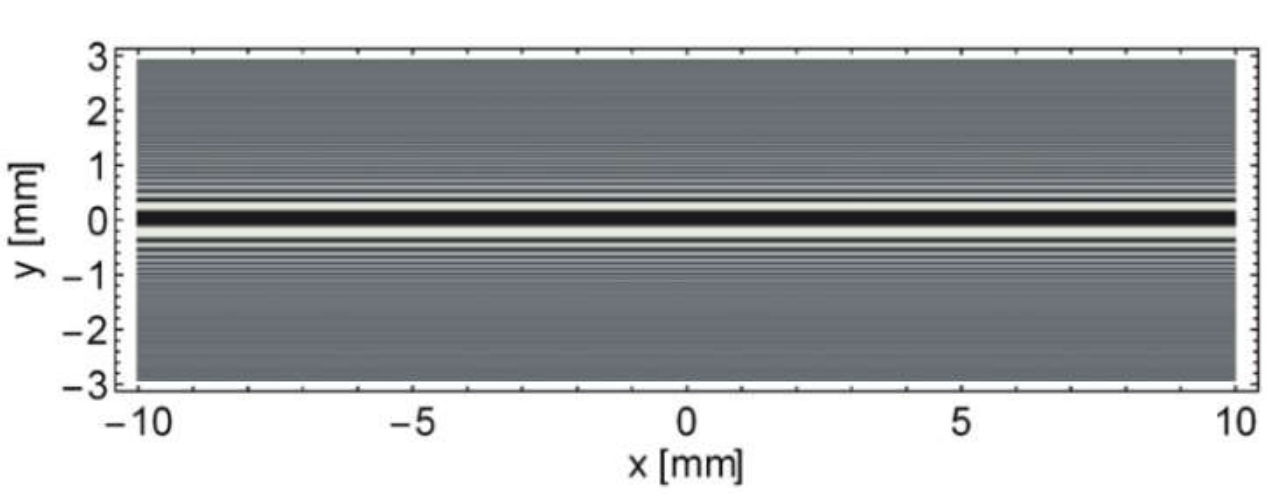

**Fig. 3.** The diffraction pattern of a plane wavefront from the edge of a phase plate with the same parameters as Fig. 2(a).

by determining the position of two different extrema, $x_1$ and $x_2$, and solving the following equation

$$x_2^2 - x_1^2 = m\lambda\left(\frac{n}{nR - nh + h} - \frac{1}{R}\right)^{-1}, \tag{9}$$

where, $m$ is the number of extrema in $x_1 - x_2$ interval.

Simulation of a diffraction pattern for $\lambda = 632.8$ mm, $h = 5$ mm, $n = 1.5, t = 1, R = 140$ mm, and $z_0 = 10$ mm is depicted in Fig. 2(a).

The normalized intensity profiles in the direction parallel to the plate edge for $y = 0$ and $y = 0.8$ mm versus $x$ and $x^2$ are shown in Figs. 2(b) and 2(c), respectively. To indicate the effect of plate position on the diffraction pattern, the profiles for $z_0 = 30$ mm are also presented in Figs. 2(b) and 2(c). The intensity profiles are completely the same at the line corresponds to the plate edge. The extrema position and the contrast of fringes change by changing $y$ and $z_0$. However, as can be seen from Fig. 2(c), those are periodic functions of $x^2$ with same period.

In order to compare with FD of plane wavefronts, a typical FD pattern using the same parameters as Fig. 2(a) but for a plane incoming wave, is presented in Fig. 3 (see Eq. (17) in Ref. [20]). Since, the phase difference is constant along the edge, the intensity is uniform in this direction.

## 2.2. Fresnel diffraction of spherical wavefront from a phase plate

In Fig. 1(b), the phase distribution on $x'y'$-plane for a point source at S, is obtained by converting $x'^2$ to $x'^2 + y'^2$ in Eq. (3), as

$$\varphi(x',y') = k\begin{cases} \frac{n}{2(nz_0+h)}(x'^2+y'^2) + \mathrm{nh} + \mathrm{z}_0 & y' < 0 \\ \frac{1}{2(z_0+h)}(x'^2+y'^2) + h + z_0 & y' > 0 \end{cases}. \tag{10}$$

By applying the Fresnel-Kirchhoff integral, the diffracted light amplitude in Fresnel approximation on $xy$-plane will be

$$\begin{aligned}\psi(x,y) = {} & KA\frac{\pi z}{\sqrt{-2i\,k}}(1+i) \\ & \times\Bigg\{ \mathrm{t}\left(\frac{nz_0+h}{nz_0+nz+h}\right) e^{i\frac{k}{2}\left(\frac{n}{nz_0+nz+h}(x^2+y^2)+2(\mathrm{n}-1)\mathrm{h}\right)} \\ & \times Erfc\left(\sqrt{\frac{-ik}{2z}\left(\frac{nz_0+h}{nz_0+nz+h}\right)}\,y\right) + \left(\frac{z_0+h}{z_0+z+h}\right) \\ & \times e^{i\frac{k}{2}\left(\frac{1}{z_0+z+h}(x^2+y^2)\right)} Erfc\left(\sqrt{\frac{ik}{2z}\left(\frac{z_0+h}{z_0+z+h}\right)}\,y\right)\Bigg\},\end{aligned} \tag{11}$$

where *Erfc* is the complementary error function defined as:

$$Erfc(\chi) = \frac{2}{\sqrt{\pi}}\int_{\chi}^{\infty} e^{-t^2}\,dt\,. \tag{12}$$

The corresponding normalized intensity, is given by dividing $\psi(x,y)\psi(x,y)^*$ by the undistributed intensity, as the following form:

$$\begin{aligned} I_n(x,y) = \frac{1}{4M^2}\Big\{ & t^2 N^2 |\beta(y)|^2 + \mathrm{M}^2 |\gamma(y)|^2 \\ & + 2t\,MN \times \mathrm{Re}\left[\beta(y)\gamma(y)\cos\phi(x,y)\right]\Big\}\,, \end{aligned} \tag{13}$$

where

$$\begin{cases} N = \dfrac{nz_0 + h}{nz_0 + nz + h} \\ M = \dfrac{z_0 + h}{R} \end{cases}, \quad \begin{cases} \beta(y) = Erfc\left(\sqrt{\dfrac{-ik}{2z}} N\, y\right) \\ \gamma(y) = Erfc\left(-\sqrt{\dfrac{-ik}{2z}} M\, y\right) \end{cases}, \qquad (14)$$

and

$$\phi(x,y) = \frac{k}{2}\left(\frac{n}{nR - nh + h} - \frac{1}{R}\right)(x^2 + y^2) + k(\mathrm{n}-1)\,\mathrm{h}\,. \qquad (15)$$

According to Eqs. (13) and (15) along to the lines parallel to the plate edge, the normalized intensity is a periodic function of $\phi\ (x,y)$ and $\phi$ is a linearly varying function of $x^2$ with the slope same as Eq. (8) that is independent to the plate position.

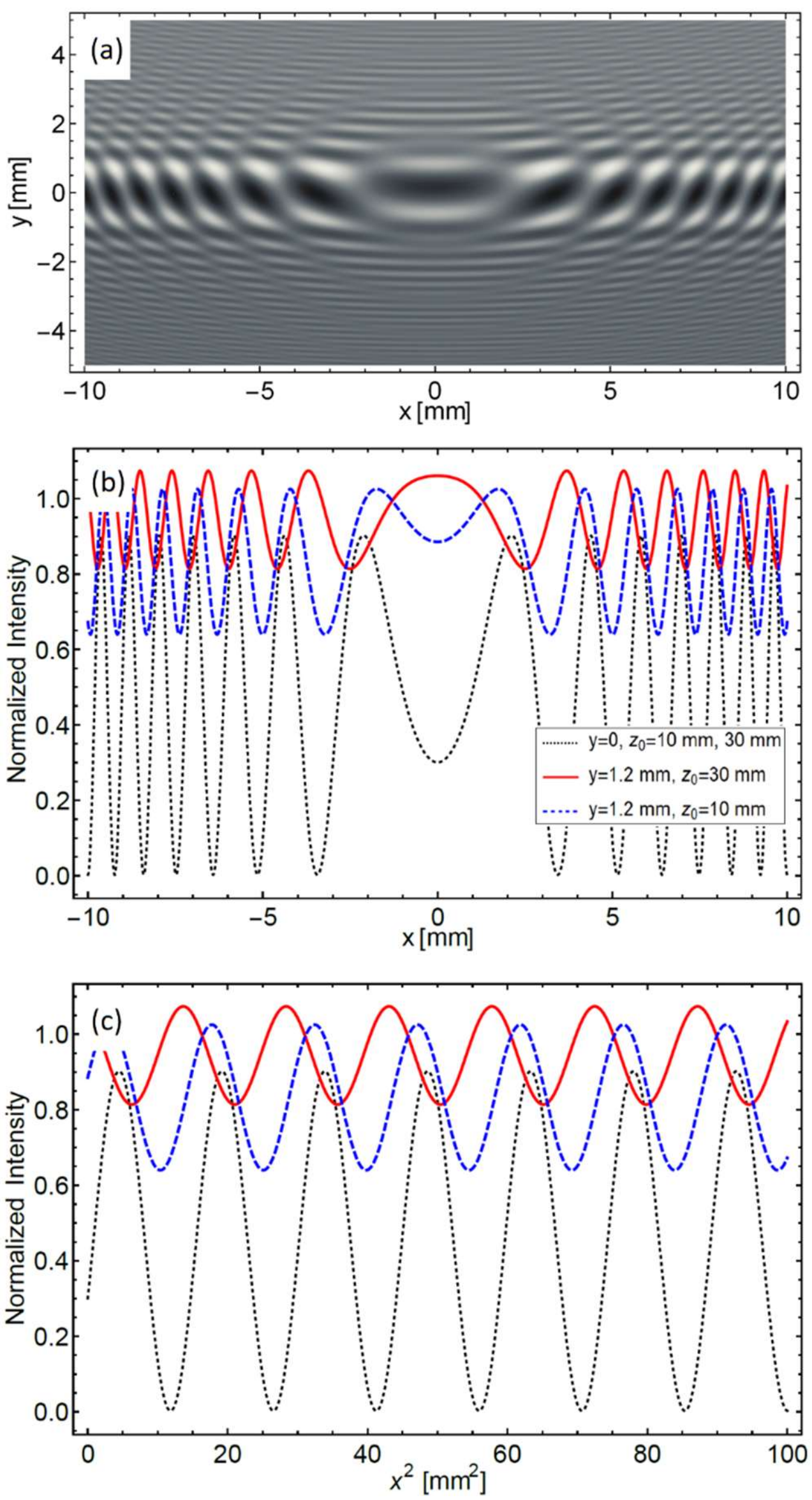

**Fig. 4.** (a) The simulated Fresnel diffraction pattern of spherical wave for the same parameters as Fig. 2(a). (b) The normalized intensity profiles in the direction parallel to the plate edge for different values of $y$ and $z_0$, and (c) the corresponding normalized intensity versus $x^2$.

Figure 4(a) shows the simulation of a spherical wavefront diffraction pattern for the same parameters as Fig. 2(a). The normalized intensity profiles in the direction parallel to the plate edge versus $x$ and $x^2$ are illustrated in Figs. 4(b) and 4(c), for different values of $y$ and $z_0$.

## 3. EXPERIMENTS AND RESULTS

In order to evaluate the simulations presented in previous Section, and measure the RWC, the following experiments were performed. The parallel beam of a He-Ne laser of wavelength 632.8 nm is focused on a slit by a cylindrical lens and then strikes the boundary of a 7.518 mm-thick (measured by the diffraction method described in Ref. [23]) plane parallel plate of BK7 with refractive index of 1.5151. The resulting diffraction patterns are recorded by a CMOS sensor of 5.19 $\mu$m pixel pitch (Canon D450) connected to a personal computer. The sensor is mounted on a linear translation stage with the position accuracy of 0.01 mm. The typical diffraction pattern obtained at a distance of about $R = 140$ mm from the focal plane is shown in Fig. 5(a). The circles and squares in Fig. 6(a) represent the phase difference of the fringes minimums from the first minimum on the right and left side of the optical axis versus $x^2$ along the plate edge ($y = 0$), respectively. The solid curve is obtained by fitting $\Delta\phi$ from Eq. (8). The corresponding RWC is 138.32 mm with 0.02% fitting uncertainty. In order to show validity of the method, the sensor is displaced by 20.00 mm along the optical axis, and another diffraction pattern is recorded (Fig. 5(b)). The corresponding experimental data and respective fitting line are represented in Fig. 6(b), which provides the value of 158.35 mm with 0.02% uncertainty for RWC. The difference between two measured RWCs is 20.03 mm that is quite consistent with the sensor displacement.

We also performed the experiment for the spherical wavefront which is obtained by passing the laser beam through a spatial filter. Figures 5(c) and 5(d) show the typical diffraction patterns obtained before and after displacing the sensor by 20.00 mm. The corresponding experimental data along the plate edge and fitting by Eq. (15) (solid lines) are shown in Figs. 7(a) and 7(b). The obtained values for RWC of spherical waves are 140.24 mm $\pm 0.03\%$ and 160.21 mm $\pm 0.03\%$. Also, the experimental data and fitted curve (solid line) for $y = 0.3$ mm of diffraction pattern of Fig. 5(d) are represented in Fig. 7(c). The slope of respective fitting line is same as Fig. 7(b) which provides the value of RWC within the same precision range.

To investigate the effect of the plate position on the diffracted intensity distribution, the diffraction pattern is obtained for plate displacement along the optical axis of about 1.5 cm towards the detector (Fig. 5(e)). Dashed curves in Figs. 7(b) and 7(c) are fitted on the corresponding experimental data for $y = 0$ and $y = 0.3$ mm, respectively. As shown in Figs. 7(b) and 7(c), the phase distributions along the plate edge ($y = 0$) are the same before and after the plate displacement. In addition, the phase distribution lines are parallel for two different $y$-values that provides the same RWC along the lines parallel to the plate edge, independent to the plate position.

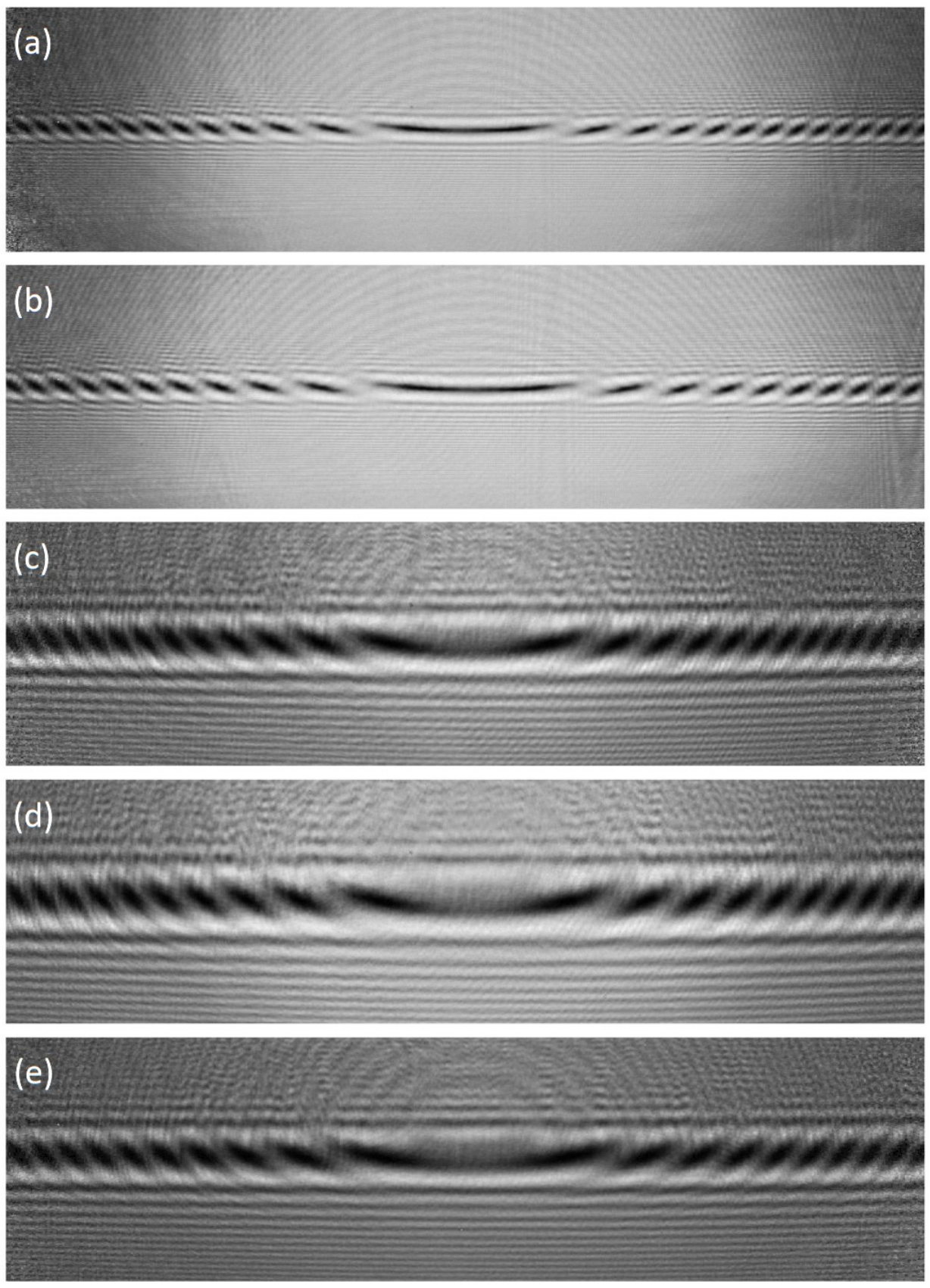

**Fig. 5.** (a), (b) Normalized experimental FD patterns obtained by illuminating a 7.518 mm-thick BK7 plate edge with a cylindrical wave of a He-Ne laser, before and after sensor displacement of 20.00 mm; (c), (d) are same as (a) and (b), but for spherical wave; (e) the diffraction pattern recorded in the same conditions as (d) when the plate is displaced along the optical axis of about 1.5 cm.

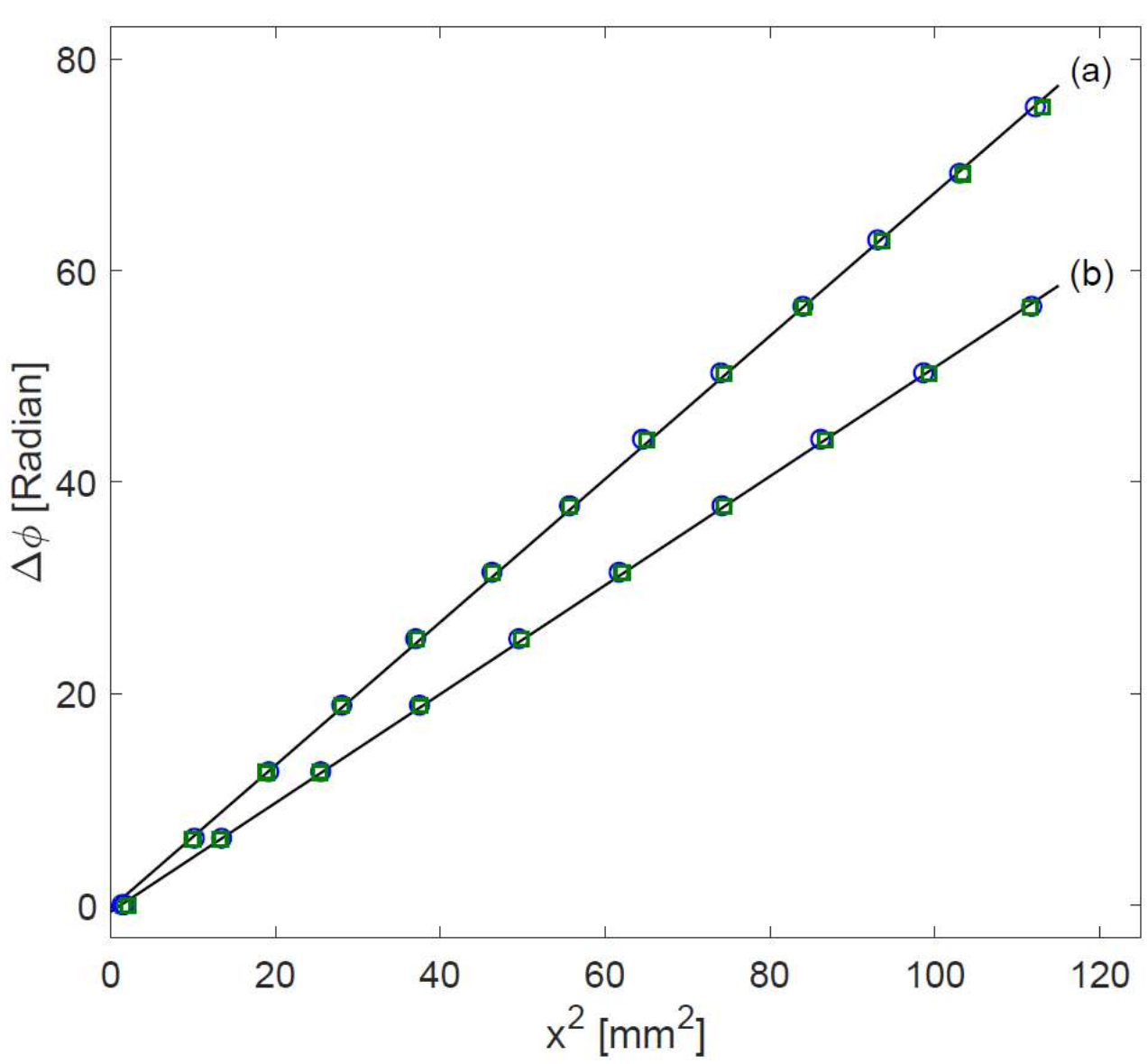

**Fig. 6.** (a) Circles and squares represent the phase difference of the cylindrical wave FD fringes minimums corresponding to Fig. 5(a) from the first minimum on the right and left side of the optical axis, along the plate edge, versus $x^2$; (b) same as (a) after displacement of the sensor along the optical axis (corresponding to Fig. 5(b)). The lines are obtained by fitting.

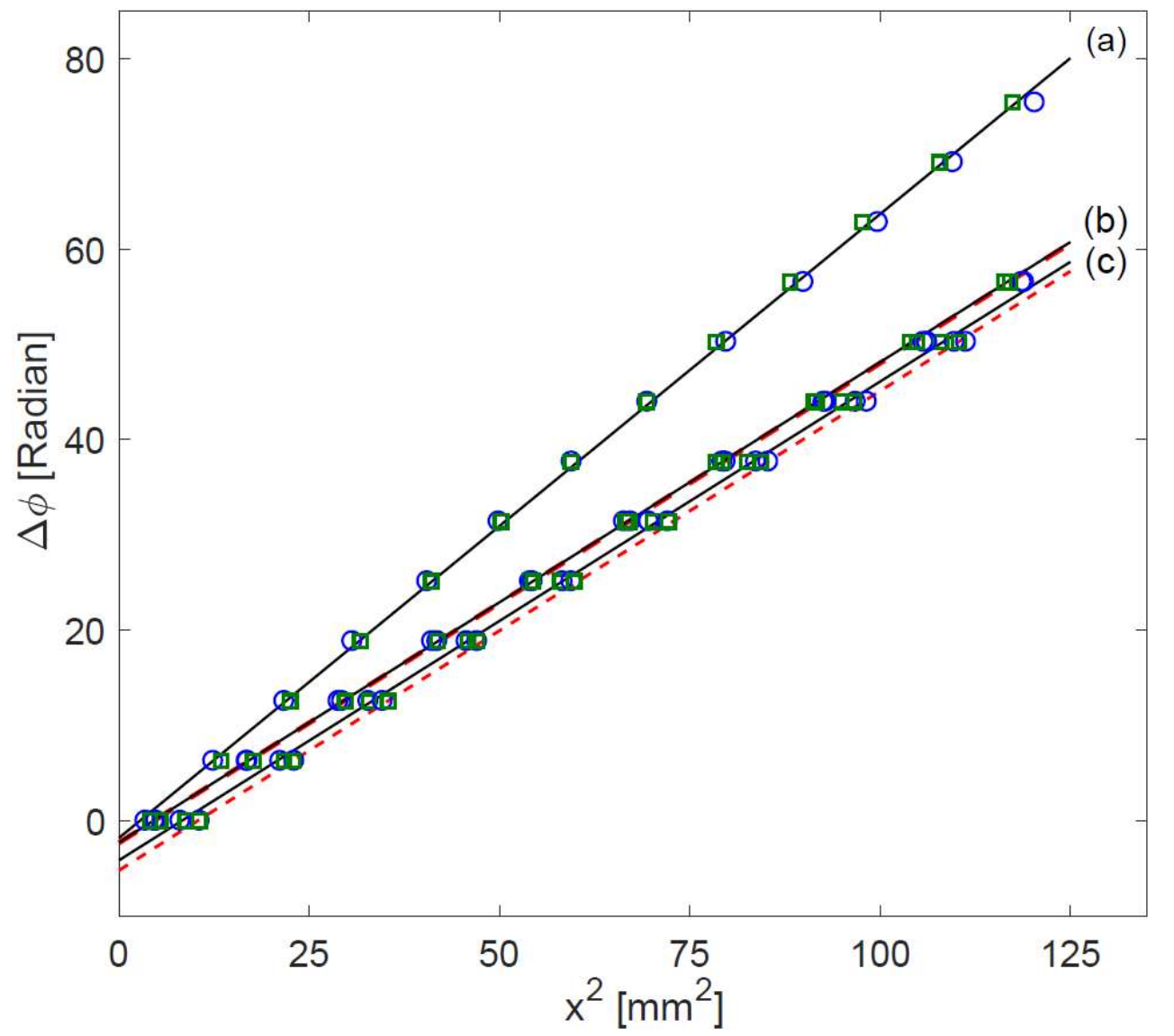

**Fig. 7.** Circles and squares represent the phase difference of the spherical wave FD fringes minimums (corresponding to Figs. 5(c)-(e)) from the first minimum on the right and left side of the optical axis versus $x^2$. The solid lines are obtained by fitting to the experimental data along the plate edge (y=0), for (a) before, and (b) after displacement of the sensor. (c) shows similar results, but for data corresponding to a line parallel to the plate edge (y=0.3 mm). Dashed lines in (b) and (c) are fitted on corresponding experimental data for y=0 and y=0.3 mm (respectively) when the plate is displaced along the optical axis.

## 4. CONCLUSIONS

Fresnel diffraction of cylindrical and spherical wavefronts from a phase step in transmission mode is formulated, which is a significant topic in physical optics. It is shown that the intensity distribution is a periodic function along the lines parallel to the plate edge. The phase distribution in this direction is a linear function of the position squared. It is also independent to the plate position that provides a straight forward approach for optical metrology such as the phase distribution, the radius of wavefront curvature (RWC), light wavelength, thickness and refractive index of the step, and displacement by a single-shot recording.

As examples of application, the phase distribution, the RWC, and displacement are determined. The method can be applied easily without using complicated accessories. Mechanical and optical noises are low in comparison with interferometric methods.